\documentclass[twocolumn,showpacs,preprintnumbers,amsmath,amssymb,superscriptaddress,a4paper]{revtex4}
\usepackage{graphicx}%
\usepackage{color}
\usepackage{todonotes}
\usepackage{hyperref}
\hypersetup{colorlinks=true,citecolor=blue}

\begin{document}

\title{Photon thermalization and a condensation phase transition in an electrically pumped semiconductor microresonator}

\author{S. Barland}
\author{P. Azam}
\author{G.~L. Lippi}
\affiliation{Universit\'e C\^ote d'Azur, CNRS, INPHYNI, 1361 route des Lucioles 06560 Valbonne, France}
\author{R.~A. Nyman}
\affiliation{Physics Department, Blackett Laboratory, Imperial College London, Prince Consort Road, SW7 2AZ, United Kingdom}
\author{R. Kaiser}
\affiliation{Universit\'e C\^ote d'Azur, CNRS, INPHYNI, 1361 route des Lucioles 06560 Valbonne, France}

\date{\today}
\newcommand{\stephane}[2][inline]{\todo[#1, color=blue!5!white]{\small \texttt{SB}: #2}}

\newcommand{\robNote}[2][inline]{\todo[#1, color=blue!5!white]{\small \texttt{Rob}: #2}}

\begin{abstract}

We report on an experimental study of photon thermalization and condensation in a semiconductor microresonator in the weak-coupling regime. We measure the dispersion relation of light and the photon mass in a single-wavelength, broad-area resonator. The observed luminescence spectrum is compatible with a room-temperature, thermal-equilibrium distribution. A phase transition, identified by a saturation of the population at high energies and a superlinear increase of the occupation at low energy, takes place when the phase-space density is of order unity. We explain our observations by Bose-Einstein condensation of photons in equilibrium with a particle reservoir and discuss the relation with laser emission.

\end{abstract}

\pacs{Valid PACS appear here}
                             
\maketitle

\newcommand{\fig}[1]{Fig.~\ref{#1}}

\makeatletter
\@input{modtime}
\makeatother

Bose-Einstein condensation (BEC) has been an important topic in both fundamental research and applied physics, with many studies involving superfluidity, superconductivity and ultracold particles.  While there particle-particle interactions are typically sufficiently strong for thermalization to occur during the experimental time scales, direct photon-photon interactions are often too small to allow for thermalization among photons to be observed. To make light reach equilibrium without destroying the optical excitations requires the introduction of both a medium which interacts strongly with the light~\cite{wurfel1982} and a gap in the optical density of states to provide an energy minimum in the optical range. Near-planar optical microresonators with mirror spacing of the order of the wavelength can provide the necessary density of states, with a characteristic paraxial parabolic form~\cite{lugiato1987,chiao1999bogoliubov,klaers2010a} for the  dispersion relation. 

In semiconductor microresonators, where light and matter coherently interact faster than dissipative processes the condensation of quasi-particles known as polaritons is well established \cite{kasprzak2006,daskalakis2014,plumhof2014}. In the highest quality samples, the polariton thermalization rate, compared to the exciton's lifetime, is high enough to show near-equilibrium BEC~\cite{sun2017}. In the absence of excitons, light and matter interact mostly incoherently (but potentially very fast) through absorption and emission processes. Photon BEC has been shown to occur in several experimental devices including fluorescent dyes at room temperature coupled to open optical microresonators~\cite{klaers2010, marelic2015, greveling2018} and plasmonic lattices~\cite{hakala2018}, with excellent agreement with a microscopic theory  based on  dissipative cavity quantum-electrodynamics~\cite{kirton2013}. Furthermore, light has been shown to thermalize~\cite{weill2017} and condense~\cite{weill2019} by incoherent interaction with erbium/ytterbium colour centres in optical fibre resonators.

Here we observe thermalization and a phase transition displaying hallmarks of photon BEC in an electrically pumped, broad-area semiconductor microresonator in the regime of weak coherent coupling but substantial incoherent coupling. This type of device is very widely used, suggesting that photon BEC could be more commonplace than widely believed.

Vertical-Cavity Surface-Emitting Lasers (VCSELs) are a mature semiconductor-based technology, where coherent optical conversion is very efficiently obtained from a heterostructure. Together with other technical advantages, a fabrication process particularly suited for mass-production enables their routine applications which range from optical telecommunication and distance sensing to very high power devices for illumination and heating  \cite{michalzik2012vcsels,moench2015high}.  These devices are typically described as light emitters where the  threshold for coherent emission is reached when optical gain compensates losses.
Broad-area multimode VCSELs have been so-far analyzed in great depth in the mean-field and semiclassical settings because of their large aspect ratio, which is essential for the analysis of transverse spatial phenomena including optical morphogenesis, localized states and vortices \cite{spinelli1998spatial,hegarty1999pattern,ackemann2000spatial,barland2002cavity,tanguy2008realization,barbay2008homoclinic,genevet2010bistable, ackemann2009fundamentals,barbay2011cavity}. In contrast to the vast literature about spatio-temporal phenomena, no detailed analysis of the spectral properties of these devices in relation to BEC has been yet reported. 
Three key properties of VCSELs make photon thermalisation and BEC almost inevitable: (i) re-absorption of light in these structures by the gain medium is strong, (ii) the semiconductor medium follows a thermal relation between gain and emission, the van Roosbroeck-Shockley relation~\cite{bhattacharya2012}, and (iii) by construction, broad-area VCSELs are longitudinally single-mode but transversely many mode, so show a low-energy cutoff in the density of states for light.

In the following we show that the dispersion relation of light in a broad-area VCSEL leads to an effective photon mass in absence of excitons. We measure a luminescence spectrum compatible with a room-temperature thermal-equilibrium distribution, which suggests that thermalization of light via interaction with a thermal reservoir occurs. This is confirmed by the phase transition observed at high photon density, which is identified both by a superlinear increase of the occupation of low energy modes and a saturation in the population of higher-energy levels as the total population is increased. At the transition, the phase-space density is of order unity, which is expected for BEC in a flat two-dimensional potential for a finite-sized system.

\begin{figure}[t]
	\setlength{\unitlength}{1cm}
	%
	%
	\includegraphics[width=0.45\textwidth]{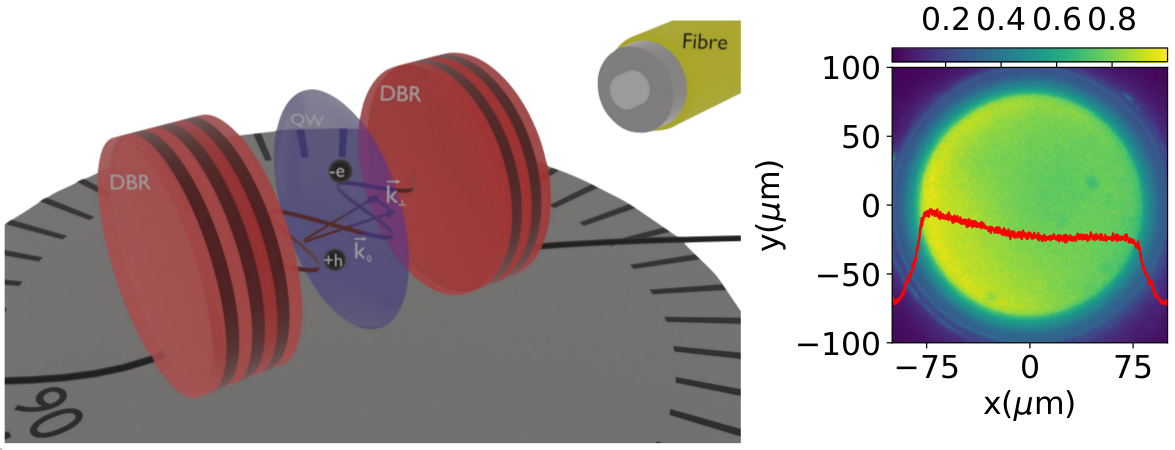}
	\caption{\label{fig:setup}Conceptual schematics of the experimental platform: light is confined in a 150~$\mu m$ diameter and 1-$\lambda$-long microresonator closed by flat high-reflectivity dielectric mirrors where an electrically pumped thin layer of quantum wells provides nonlinearity (see \cite{grabherr1998bottom} for a detailed description of the device). After 7~cm free space propagation, light is collected by a multimode optical fiber. Right: color-coded image of light emitted from the front-facing mirror, the red line is a cut along the horizontal diameter.}
\end{figure}

The experimental device (\fig{fig:setup}, left) consists of a thin layer of quantum-well active material, where the light-matter interaction takes place, enclosed in a microresonator formed by high-reflectivity dielectric mirrors. It is fabricated as an electrically excited, 150~$\mu$m diameter VCSEL. The material consists of three InGaAs quantum wells embedded in GaAs barriers, themselves surrounded by AlGaAs cladding which serves as spacing to obtain a single-wavelength thickness of the inner resonator. The back and front mirrors are composed of respectively 30 and 24 pairs of Al$_{\rm 0.9}$Ga$_{\rm 0.1}$As--GaAs \cite{grabherr1998bottom,ackemann2000spatial} with a cavity resonance close to 975~nm. Photons are emitted inside the resonator thanks to electron--hole recombinations in the quantum-well material, which was prepared to have a photoluminescence maximum between 980 and 1000~nm under operating conditions, thereby matching the detuning requirements outlined for photon BEC in~\cite{kirton2013}. The flat mirrors are expected to lead to uniform distributions of electric field and density of states, but current diffusion and insulating barriers lead to a slightly higher rate of photon emission close to the edges due to current crowding \cite{michalzik1998high} (right panel of \fig{fig:setup}). The temperature of the device's substrate is actively stabilized at 293~K.  Excitons are not formed significantly in GaAs at room temperature and the experiment is performed in the weak-coupling regime.

We analyze the spectral light properties as a function of emission angle with adequate numerical aperture using a goniometric setup. The semiconductor micro-resonator is places at the center of a 7-cm diameter programmable rotation stage. No collimating or focusing optics are used and light is collected by a 200~$\mu$m-core (multimode) optical fiber placed about 7~cm from the cavity, ensuring far-field detection (Fraunhofer distance $\approx$ 4~cm).  The angular spectral analysis is obtained by rotating the sample around the stage's axis while keeping the optical fiber position and alignment stationary and fixed with respect to the center of the micro-resonator.

\begin{figure}[t]
	\setlength{\unitlength}{1cm}
	\includegraphics[width=0.45\textwidth]{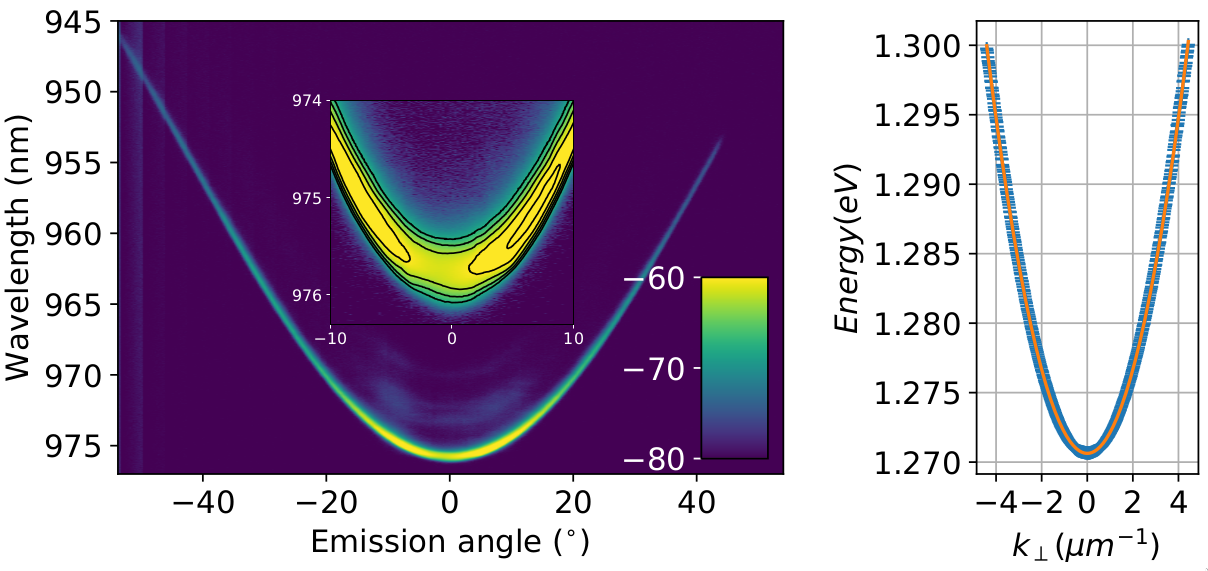}
	\caption{\label{fig:parabola}Experimental measurement of the dispersion relation. Left: color-coded emission spectra in~dBm with angular resolution (pumping current 150~mA, near threshold), the inset is a zoom on the central part. Right: corresponding dispersion relation. The blue crosses are the experimental data and the orange continuous line is a parabolic fit.}
\end{figure}

The measurement angle is varied over a range of 107.8$^{\circ}$ in steps of 0.2$^{\circ}$, matching the resolution limit set by the fiber diameter and propagation distance. For each angle we measure an optical spectrum in the interval 940.0~nm $\le \lambda \le$ 980.7~nm (i.e. 1.264--1.319~eV), as shown in \fig{fig:parabola}. The color-coded spectra (resolution 0.1~nm, left panel) are shown as a function of the emission angle, the total image containing more than $5.5\times 10^6$ data points. The short and planar resonator, with its large aspect ratio allows us to clearly observe the parabolic shape expected in a Fabry-Perot cavity for the transverse component of the light's dispersion relation~\cite{lugiato1987,chiao1999bogoliubov} which is verified (right panel \fig{fig:parabola}) over a remarkably large emission angle ($\approx$ 86$^{\circ}$).  We attribute the asymmetry appearing at large emission angles between the two sides to tiny misalignments of the mirrors inherent to the growth process and to inhomogeneous current distribution (right panel, \fig{fig:setup}).

By fitting the energy-momentum relation $E(p_{\perp})~=~m_{ph}c^2 + p_{\perp}^2/2m_{ph}$, an effective mass $m_{ph}$ for the photon can be estimated either from the offset (cut-off energy) or from the curvature term. Here, $c$ is the speed of light in the medium and $p_{\perp}$ the in-plane momentum of the light. From the latter, we find a photon mass $m_{ph}=(2.2881 \pm 0.0008) \times 10^{-35}$~kg indicating an excellent fit. This measurement has been repeated over 43 values of pumping current between 20 and 200~mA and the mass was found to be remarkably constant $(2.28 \le m_{ph} \le  2.31) \times 10^{-35}$~kg.
The cut-off energy of the cavity yields an alternative method to extract the effective photon mass. Using the index of refraction as a free parameter, we obtain the same effective photon mass for a index of refraction of 3.2, consistent with the material properties of our device~\footnote{At $\lambda =$ 1~$\mu m$, $n = 3.5; 3.0; 3.0$ for GaAs, Al$_{0.9}$Ga$_{0.1}$As and AlGaAs, respectively.}. If strong coherent exciton-photon interactions were present the effective masses from the two methods would differ. Here they are identical, which confirms that (as expected at room temperature and for this construction) our device works with photons and not polaritons~\cite{klaers2010,marelic2016phase}.

\begin{figure}[t]
	\setlength{\unitlength}{1cm}
	\includegraphics[width=0.42\textwidth]{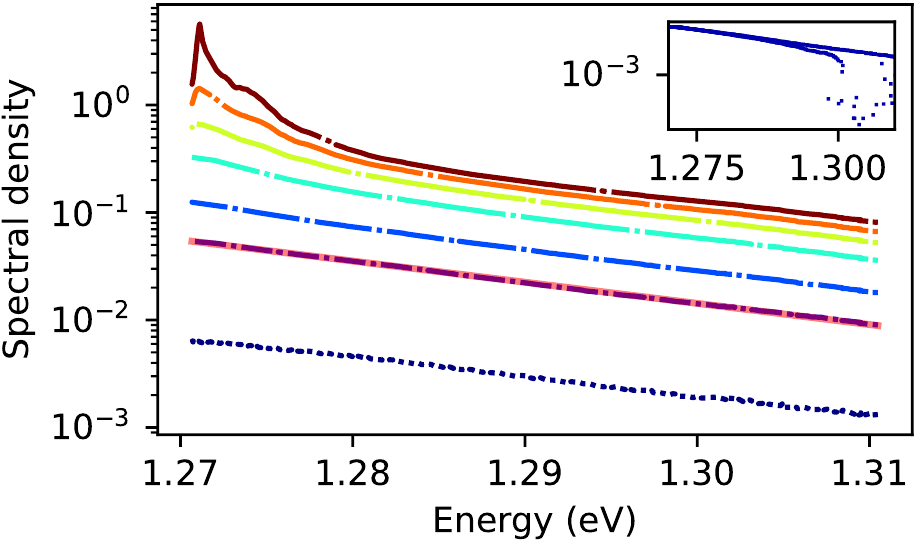}
	\caption{\label{fig:spectraldensity}Spectral density for different intracavity photon numbers. The data were acquired, from dotted dark blue (bottom) to dot- long-dashed brown (top), with pumping current values of: 4.05, 24.33, 44.60, 85.15, 125.70, 166.25, 206.79~mA respectively. The pink continuous line is an exponential fit which gives a temperature of 256~K. Inset: the asymmetry visible in \fig{fig:parabola} materializes as a cut-off in the spectral density on one side.}
\end{figure}

With this established dispersion relation, we can look at the distribution of light over all the available energy band. Although this is in principle a purely spectral measurement it cannot be realized simply with an optical spectrum analyser due to the extremely large numerical aperture which is required. Instead we rely on the previously determined dispersion relation and perform the measurements in the angular domain.  The emitted power was measured over an approximately symmetric range of 107.8~$^\circ$ by steps of 0.2~$^\circ$ for 62 different pumping current values and a sample of the results is shown on \fig{fig:spectraldensity}. The data, acquired at constant solid angle of collection, has been scaled to correspond to a constant range in $k_{\perp}$. For the lowest pump values (the lowest three traces), the distributions fit very well a Boltzmann distribution with a constant density of states, as expected in a homogeneous two-dimensional microresonator. In \fig{fig:spectraldensity}, the continuous line is the Boltzmann law fit of the experimentally observed distribution at pumping current of 24.3~mA and it leads to a temperature of 256~K. Similar values are found for different pumping levels. This observation is crucial since the radiation temperature defined in \cite{wurfel1982} is expected to be identical to that of the medium. Although we have at present no explanation for this slight underestimation of the actual temperature, it appears to be common for temperatures estimated from light emission spectra \cite{walker2018driven,klaers2010}. Unlike atomic systems, semiconductor media have a transparency current density below which absorption dominates over amplification. We did not observe critically different curves in our data, which may in fact all be taken above this value since it can only be roughly estimated from material tables and geometry \cite{coldren} to be about 8~mA.

Assuming equal probability of absorption and stimulated emission, based on tabulated values \cite{coldren} for the active region and on geometrical parameters of our device, the intraband thermalization time can be estimated to be of the order of few tens of femtoseconds. This time is extremely short compared to the photon lifetime in the resonator $\tau$, which can be estimated from the resonator length and mirror reflectivities to be between $2.5 \leq \tau \leq 10$~ps depending on the absorption in the spacing layers and mirrors \cite{michalzik2003operating}. 
We explain the observed spectral distribution with the corresponding temperature matching the expected value without any free parameter in terms of equilibrium thermalization of radiation.

At higher photon density a deviation from the exponential distribution emerges close to the low-energy cut-off, while the distribution in the high-energy region remains exponential. For the two largest pumping currents (top two curves) the maximal density is not strictly in the lowest energy mode but at 1.2711~eV instead of 1.2707~eV. First we emphasize that this observation is only possible because of the excellent resolution allowed by angular domain measurements in a goniometric setup, since \AA{}ngstrom resolution together with extremely large numerical aperture are required. We attribute this tiny deviation from the exact $k_{\perp}=0$ mode to spatial phenomena caused by inhomogeneities of the device. Indeed there is a region of higher photon density close to the boundaries of the active layer due to current spread and insulating boundaries (\fig{fig:setup} left and \cite{ackemann2000spatial}) and furthermore, the temperature profile of the device may not be constant along the radial direction due to Joule heating.

Despite a small asymmetry at high energy (\fig{fig:spectraldensity} inset and \fig{fig:parabola}), the deviation from the Boltzmann distribution at low energy is clear for both signs of $k_{\perp}$ and indicates the existence of a phase transition at higher values of the pumping parameter.

\begin{figure}[t]
	\setlength{\unitlength}{1cm}
	\begin{picture}(8,5)
		\put(0,0.){\includegraphics[width=0.42\textwidth]{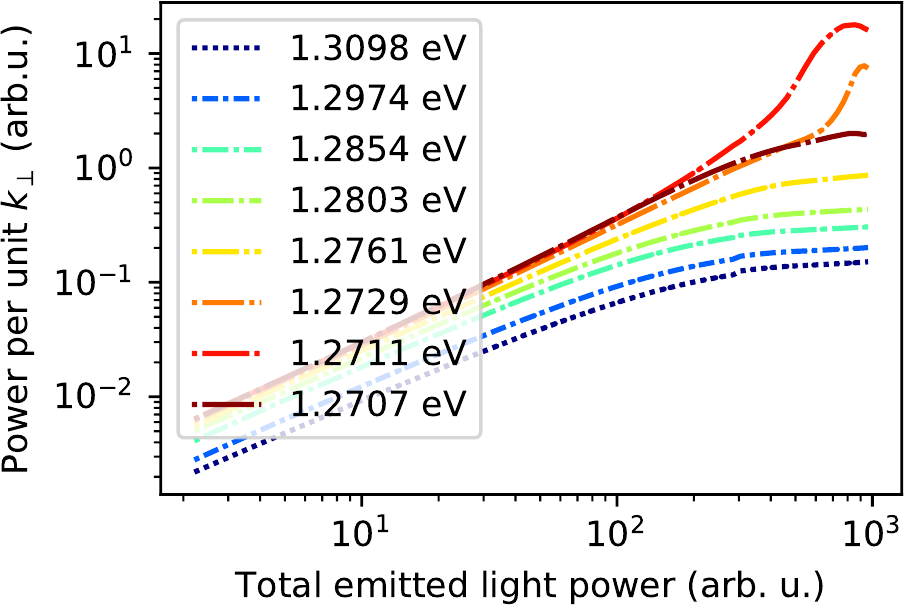}}
       		\put(4.8,0.50){\includegraphics[width=3.15cm]{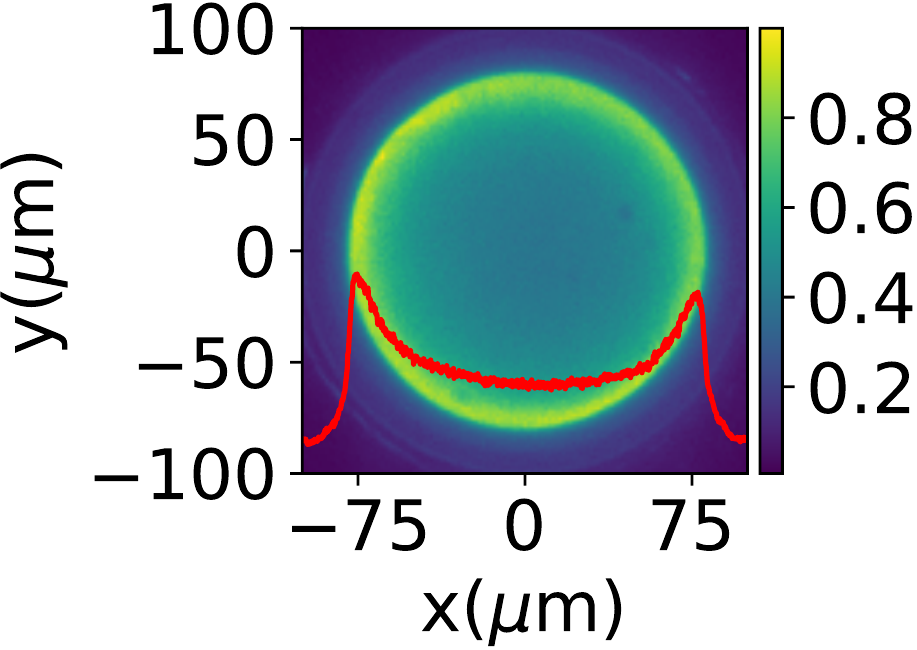}}
	\end{picture}
	\caption{\label{fig:transition} Spectrally resolved light intensity as a function of the total photon number. The population excess at low energies is accompanied by a saturation in the population of the high-energy states. Inset: The image of the quantum-well region shows a much higher photon density close to the edges, very visible along the horizontal cut (continuous line).}
\end{figure}

To characterise this phase transition we plot the same spectrally resolved light intensity as a function of the total emitted light power, proportional to the intracavity photon number (\fig{fig:transition}, left). We show  data for eight values of $\omega$ which correspond to emission angles of 53.2, 41.2, 29.2, 17.2, 5.2 and 0.2~$^{\circ}$ respectively from the bottom to top curves.
At low total light power, the emitted light power per unit $k_{\perp}$ scales linearly with the total emitted power. Beyond 3mW deviations appear. The bottom three curves (highest energy) display a clear saturation (sub-linear growth with respect to total power) while the top curves demonstrate excess population (superlinear) in the lower-energy states. Both the superlinear and the saturation are also visible when the injection current is used as horizontal axis. In this case, the total emitted power also shows the typical laser characteristic (see supplemental information). The saturation is indicative that the gain clamping associated with the stimulated emission into the ground state causes a clamping of gain for all modes. 

Both the population excess at low energy and the saturation of growth at high energy are consistent with a second-order phase transition in a finite-size system in which the whole spectrum of radiation shares the same reservoir.  The inset of \fig{fig:transition} shows that the most populated mode is essentially a ring close to the edge, most probably due to a higher current density in this region \cite{ackemann2000spatial}. It is striking that intraband thermalization of carriers is still an effective coupling mechanism even in presence of spatial inhomogeneity. 

To compare our observations to previous realizations of photon BEC in two-dimensional systems \cite{damm2017first}, we first compute the thermal de Broglie wavelength $\Lambda_{th}=h/\sqrt{2\pi m k_B T}$, which for our parameters  ($m=2.28~10^{-35}kg$, $T\approx256K$) yields $\Lambda_{th}\approx 0.9\mu m$. The areal photon density $\rho_{2D}$ at the transition can be computed from the emitted power at threshold ($P_0=3~mW$), the device diameter ($\Phi=150~\mu$m) and the resonator lifetime $\tau$, estimated to be between 2.5 and 10~ps, $\rho_{2D}=\frac{4P_0\tau}{\hbar\omega\pi\Phi^2}$. With these numbers, we find that the photon phase space density at threshold is of order unity with $1.7 \leq \rho_{2D}\Lambda_{th}^2 \leq 7.1$.
Given the uncertainties of the parameters of the resonator, the threshold behaviour observed in  \fig{fig:transition} is thus consistent with a Bose-Einstein type transition.

The observations reported above are compatible with a second-order phase transition leading to a macroscopic occupation of the lowest energy band. We underline that  many open systems can exhibit a similar macroscopic occupation of a single `quantum' state, without relying on a thermalization process. This is for instance the case for most lasers. In the present case there is no doubt that the observed equilibrium distribution is not mediated by conservative  interparticle collisions, which would preserve at any time the total number of particles and would be a pure Kerr nonlinearity. Instead,  we attribute thermalization, and subsequent macroscopic occupation of a single state, to  particle exchange with a thermal reservoir, a process which involves all possible interactions between light and matter, including absorption and spontaneous or stimulated emission. 
We also note that in our device, no parabolic confinment potential has been engineered. In an extremely large and perfectly homogeneous device, one would thus not expect BEC, but rather the appearance of a state showing either Berezinskii-Kosterlitz-Thouless (BKT) or Kardar-Parisi-Zhang (KPZ) correlations, depending on the strength of the drive and dissipation \cite{altman2015two}. In our device, finite-size effects are known to be negligible in a strongly nonlinear regime \cite{ackemann2009fundamentals,barbay2011cavity} but residual deviations from perfect homogeneity clearly remain relevant. Further insights could be obtained via amplitude and intensity correlation measurements  \cite{schmitt2014observation}.

In summary, we have analyzed the spectral properties of an electrically pumped broad-area VCSEL and found that a Boltzmann law adequately fits spectral density over a broad range of parameters, leading to a temperature compatible with the experimental conditions. The accumulation of light in the lower-energy states when a critical intracavity photon number is reached is accompanied by saturation in the populations of high-energy states. This confirms that light thermalization through interaction with the semiconductor medium occurs throughout this phase transition which is well described by Bose-Einstein statistics. At the transition the phase-space density is of order unity which is compatible with an interpretation in terms of photon BEC. These observations have been realized in a semiconductor microresonator whose only particularities with respect to the many VCSELs currently in use is a very large aspect ratio (which sets the adequate dispersion relation over a broad spectral range) and a specific sign of detuning between resonator and luminescence maximum. This suggests that photon BEC could be much more commonplace than initially envisioned and that semiconductor microresonators in the weak-coupling regime are easily manufacturable sources of thermalized light, where spectral properties and phase space density can be further engineered. On a more fundamental level, these devices emerge as an extremely versatile and convenient platform for the exploration of photonic phase transitions. Specifically, spatially resolved amplitude, phase and intensity correlation measurements could confirm the grand canonical conditions of the experiment and explore the possibilities of BKT or KPZ transitions.
This experiment should also inspire further theoretical investigation. First the role of finite boundary conditions in a planar geometry must be carefully addressed. Second, unlike all other systems achieving light thermalization, here the medium is not made of a finite number of discrete colour centres, but, rather, of electron and hole excitations in variable numbers, thus rendering not directly applicable existing theories of photon thermalization.  

\begin{acknowledgments}
{This  work  has  been  supported  by  the European Union's Horizon 2020 research and innovation programme under grant agreement No 820392 in the PhoQuS project framework, and by EPSRC(UK) under grant EP/S000755/1.}
\end{acknowledgments}

\bibliography{thermal}

\end{document}